# The General Solution to Vlasov Equation and Linear Landau Damping

Deng Zhou
Institute of Plasma Physics, Chinese Academy of Sciences
Hefei 230031, P. R. China

ABSTRACT

A general solution to linearized Vlasov equation for an electron electrostatic wave in a homogeneous unmagnetized plasma is derived. The quasi-linear diffusion coefficient resulting from this solution is a continuous function of $\omega$ at $Im(\omega) = 0$ in contrast to that derived from the traditional Vlasov treatment. The general solution is also equivalent to the Landau's treatment of the plasma normal oscillations, and hence leads to the well-known Landau damping.

Linear electron plasma waves in a collisionless plasma can be obtained by solving the linearized Vlasov equation together with the Poisson equation, which was first treated by Vlasov[1]. The dispersion relation is given by

$$D(\omega, k) = 1 + \frac{e^2}{mk\epsilon_0} \int_{-\infty}^{+\infty} \frac{\partial f_0/\partial v}{\omega - kv} dv = 0 \quad (1)$$

where $\omega$ and $k$ are respectively the frequency and the wave number, other symbols are obvious. $D(\omega, k)$ is usually called the plasma dielectric function. From Eq. (1), we can get the usual Langmuir wave. However, it is inadequate to treat all the effects of thermal particles. We notice from Eq. (1) that there is a singularity at $v = \omega/k$ in the integral and $D(\omega, k)$ is not a continuous function of $\omega$ at $Im(\omega) = 0$.

To overcome these insufficiencies, Landau solved the Vlasov equation as an initial-value problem and introduced the deformed integration route in the plasma dielectric function to make sure that $D(\omega, k)$ is an analytical function in the whole complex plane of $\omega$. Landau's treatment leads to the famous landau damping[2], which was later demonstrated in experiments[3]. The introduction on the two kinetic treatments of plasma waves can be found in most plasma textbooks, such as Ref.[4]. An overview by Ryutov summarized the studies of Landau damping by the end of twentieth century[5]. The Landau damping derived by Landau is a pure achievement of applied mathematics. On it's physical explanation different people have different points of view ( see, for example, Drummond[6] and references therein ).

On a weak nonlinear level, a quasi-linear approach is adopted to describe the interaction between waves and particles[7,8]. The averaged background particle distribution may experience diffusion in the velocity space due to the wave-particle interaction. Taking the one dimensional electrostatic case as an example, one gets distribution evolution equation

$$\frac{\partial f_0}{\partial t} = \frac{\partial}{\partial v} \mathcal{D} \frac{\partial f_0}{\partial v} \quad (2)$$

with the quasi-linear diffusion coefficient

$$\mathcal{D} \propto \frac{\gamma}{(\Omega - kv)^2 + \gamma^2} \quad (3)$$

where $\Omega(\gamma)$ is the real (imaginary) component of $\omega$. Eq. (3) is derived from the Vlasov's solution for unstable modes, i. e. $\gamma > 0$. When the resonance broadening disappears one meets a difficulty in extending (3) to the $\gamma < 0$ cases. We note that in the limit $\gamma \to 0$, Eq. (3) reduces to

$$\mathcal{D} \propto sign(\gamma)\pi\delta(\Omega - kv) \quad (4)$$

where $sign(\gamma)$ denotes sign of $\gamma$. The resonance broadening disappears and a discontinuity appears at $\gamma = 0$. To overcome this problem, one has to turn to Landau's treatment to get the distribution function. It is difficult to get an explicit function of distribution since an inverse Laplace transform is involved. To circumvent this problem[9], Hellinger *et al* recently extended the linear solution to Vlasov equation for the $\gamma < 0$ case by adding a term involving a complex Dirac delta function. However, these authors didn't realize that the extended solution is the real solution to Vlasov equation and is related to linear Landau damping, and moreover Dirac delta function is not a well-defined function in complex plane.

In the following, we derive a general solution to Vlasov equation for the electrostatic Langmuir waves and from this solution the quasi-linear coefficient is derived and shown to be continuous from growing to damped modes. We also show that the general solution is equivalent to the Landau treatment and leads to the famous landau damping.

We consider the one dimentional electrostatic oscillation in a homogeneous unmagnetized plasma, the linearized Vlasov equation is

$$\frac{\partial f_1}{\partial t} + v\frac{\partial f_1}{\partial x} = \frac{eE_1}{m}\frac{\partial f_0}{\partial v} \quad (5)$$

The perturbation of electric field and particle distribution is written in the normal mode form

$$E_1 = \hat{E}_k e^{-i(\omega t - kx)} \quad (6)$$
$$f_1 = \hat{f}_k e^{-i(\omega t - kx)} \quad (7)$$

Eq. (6) is then reduced to

$$-i\omega\hat{f}_k + ikv\hat{f}_k = \frac{e\hat{E}_k}{m}\frac{\partial f_0}{\partial v} \quad (8)$$

The general form of solution to Eq. (8) is

$$\hat{f}_k = -i\frac{e\hat{E}_k}{mk}\frac{\partial f_0}{\partial v}\left[\frac{1}{v - \omega/k} + S_{\omega,k}(v)\Delta(v, \omega/k)\right] \quad (9)$$

where $S_{\omega,k}(v)$ is any function of $\omega$, $k$ and $v$, and

$$\Delta(v, \omega/k) = \delta(v - \Omega/k) + (-i\gamma/k)\delta'(v - \Omega/k) + \frac{(-i\gamma/k)^2}{2!}\delta''(v - \Omega/k) + \cdots$$
(10)

Setting $S_{\omega,k}(v) = 0$, we recover the Vlasov treatment and get the dispersion relation Eq. (1). To prove that Eq. (9) is the solution of Eq. (8), one needs to demonstrate

$$(v - \omega/k)\Delta(v, \omega/k) = 0 \quad (11)$$

This relation is obvious for $\gamma = 0$. For $\gamma \neq 0$, we have the following integrals for any smooth function $F(v)$

$$\int_{-\infty}^{+\infty} F(v)(v - \omega/k)\delta(v, \Omega/k)dv = -(i\gamma/k)F(\Omega/k) \quad (12a)$$

$$\int_{-\infty}^{+\infty} F(v)(v - \omega/k)(-i\gamma/k)\delta'(v - \Omega/k)dv = (i\gamma/k)F(\Omega/k) - (i\gamma/k)^2 F'(\Omega/k)$$
(12b)

$$\int_{-\infty}^{+\infty} \frac{1}{n!}F(v)(v - \omega/k)(-i\gamma/k)^n\delta^{(n)}(v - \Omega/k)dv =$$
$$\frac{(i\gamma/k)^n}{(n-1)!}F^{(n-1)}(\Omega/k) - \frac{(i\gamma/k)^{n+1}}{n!}F^{(n)}(\Omega/k) \quad (12c)$$

Combining (10) (12a-c), we obtain

$$\int_{-\infty}^{+\infty} F(v)(v - \omega/k)\Delta(v, \omega/k)dv = 0 \quad (13)$$

Since $F(v)$ can be any smooth function, then from the basic theorem of function theory, Eq. (11) is verified.

The form of $\Delta(v, \omega/k)$ is superficially the Taylor expansion of the complex Dirac delta function $\delta(v - \omega/k)$. However, it is impossible to define a proper complex Dirac delta function $\delta(z)$ which is analytical at $z = 0$ since we know that a complex function analytical at whole complex plane is only a constant.

Now we need to determine $S_{\omega,k}(v)$. Without loss of generality, we set $S_{\omega,k}(v) = 0$ for $\gamma > 0$ and derive the value for $\gamma \leq 0$ through the continuity of density perturbation at $\gamma = 0$.

The density perturbation is given by

$$\hat{n}_k = -i\int_{-\infty}^{+\infty} \frac{e\hat{E}_k}{mk}\frac{\partial f_0}{\partial v}\left[\frac{1}{v - \omega/k} + S_{\omega,k}(v)\Delta(v, \omega/k)\right]dv \quad (14)$$

The integration route is from $-\infty$ to $+\infty$ in $v$-space. In Eq. (14), the principal value of integral is taken if a singularity lies on the integration route.

For $\gamma > 0$, we get from (14)

$$\hat{n}_k = i\int_{C_R} \frac{e\hat{E}_k}{mk}\frac{\partial f_0}{\partial v}\frac{1}{v - \omega/k}dv + 2\pi \frac{e\hat{E}_k}{mk}\frac{\partial f_0}{\partial v}\bigg|_{v=\omega/k} \quad (15)$$

where $C_R$ denotes the integration route of the semi-circle on the upper half complex plane from $+\infty$ to $-\infty$, as indicated in Fig. 1. For $\gamma < 0$, the density perturbation is given by

$$\hat{n}_k = i\int_{C_R} \frac{e\hat{E}_k}{mk}\frac{\partial f_0}{\partial v}\frac{1}{v-\omega/k}dv - iS\frac{e\hat{E}_k}{mk}\left[\frac{\partial f_0}{\partial v}\bigg|_{v=\frac{\omega}{k}} + (i\gamma)\frac{\partial^2 f_0}{\partial v^2}\bigg|_{v=\frac{\omega}{k}} + \frac{(i\gamma)^2}{2!}\frac{\partial^3 f_0}{\partial v^3}\bigg|_{v=\frac{\omega}{k}} + \cdots\right] \quad (16)$$

One obtains $S = 2\pi i$ for $\gamma < 0$ from (15) and (16) by the requirement of $\hat{n}_k(\gamma \to 0+) = \hat{n}_k(\gamma \to 0-)$. The same procedure yields $S = \pi i$ for $\gamma = 0$. In summary, we get the general solution to Eq. (8) given by Eq. (9) and (10) with

$$S_{\omega,k}(v) = \begin{cases} 0 & \gamma > 0 \\ \pi i & \gamma = 0 \\ 2\pi i & \gamma < 0 \end{cases} \quad (17)$$

In the weak non-linear level the averaged background particle distribution changes slowly due to the wave-particle interaction, which is described through a quasi-linear approach. Keeping to the second order of perturbation in Vlasov equation and taking a space averaging in large space scale, one obtains the distribution evolution

$$\frac{\partial f_0}{\partial t} = Re \left\langle \frac{eE_1^*}{m}\frac{\partial f_1}{\partial v}\right\rangle \quad (18)$$

where $Re$ ( $Im$ ) denotes the real ( imaginary ), bracket denotes space averaging, and the superscript star denotes complex conjugate.

Inserting (6) (7) and (9) into (18), we get the diffusion equation, Eq. (2), for distribution $f_0$ with the diffusion coefficient

$$\mathcal{D} = \left(\frac{e}{m}\right)^2 |\hat{E}_k|^2 Im\left[\frac{1}{kv-\omega} + \frac{S_{\omega,k}(v)\Delta(v,\omega/k)}{k}\right]$$
$$= \left(\frac{e}{m}\right)^2 |\hat{E}_k|^2 \left\{\frac{\gamma}{(\Omega-kv)^2+\gamma^2} + \frac{S}{ik}\left[\delta(v - \Omega/k) - \left(\frac{\gamma}{k}\right)^2 \delta''(v - \Omega/k) + \cdots\right]\right\} \quad (19)$$

For simplicity we have kept only one wave number while the real diffusion is contributed from all the excited modes.

It is obvious that the diffusion coefficient given by (19) is continuous as $\gamma \to 0\pm$, which is

$$\mathcal{D}(\gamma \to 0+) = \mathcal{D}(\gamma \to 0-) \propto \pi\delta(kv - \Omega) \quad (20)$$

The general solution, Eq. (9), can lead to linear Landau damping. Substituting the perturbation of the electric field and the distribution into the one dimensional Poisson equation

$$\epsilon_0 \frac{\partial E_1}{\partial x} = -e \int_{-\infty}^{+\infty} f_1 \, dv \quad (21)$$

we obtain the dispersion relation

$$1 - \frac{\omega_{pe}^2}{n_0 k^2} \int_{-\infty}^{+\infty} \frac{\partial f_0}{\partial v} \left[ \frac{1}{v - \omega/k} + S_{\omega,k}(v) \Delta(v, \omega/k) \right] dv \quad (22)$$

where $\omega_{pe} = \sqrt{\frac{n_0 e^2}{\epsilon_0 m}}$ is the usual Langmuir frequency and $n_0$ is the background plasma density.

Substituting (10) in (22) yields

$$1 - \frac{\omega_{pe}^2}{n_0 k^2} \int_{-\infty}^{+\infty} \frac{\partial f_0}{\partial v} \frac{1}{v - \omega/k} dv - S \frac{\omega_{pe}^2}{n_0 k^2} \frac{\partial f_0}{\partial v} \bigg|_{v = \frac{\omega}{k}} = 0 \quad (23)$$

This is exactly the plasma dielectric function one obtains from Landau's treatment by adopting a modified Landau integration contour in complex $v$-plane. Hence, the general solution to Vlasov equation is equivalent to the Landau's solution in the treatment of plasma normal modes and leads to the well-known Landau damping.

In a recent work[10], Wesson reexamined the problem of Landau damping. He solved Vlasov equation separately for $\gamma > 0$ and $\gamma < 0$ using real variables. The two cases are related through the continuity of density perturbation at $\gamma \to 0$. Our present treatment is equivalent to Wesson's since we also solve Vlasov equation separately for $\gamma > 0$ and $\gamma < 0$ and also use the continuity condition to relate the two cases. To show this, we take the case $\gamma < 0$ as an example. Like Wesson's treatment, the density perturbation is separated into two parts, $n_1^b$ and $n_1^w$. $n_1^b$ is the contribution from the basic thermal distribution, which is treated as a Dirac delta function, i. e. $f_0 = n_0 \delta(v)$. Hence

$$\begin{aligned}
n_1^b &= -i \int_{-\infty}^{+\infty} \frac{e \hat{E}_k}{mk} \frac{\partial f_0}{\partial v} \frac{1}{v - \omega/k} dv \\
&= -i \int_{-\infty}^{+\infty} \frac{e \hat{E}_k}{mk} \frac{\partial f_0}{\partial v} \frac{v - \Omega/k + i\gamma/k}{(v - \Omega/k)^2 + (\gamma/k)^2} dv \\
&= -i \frac{e \hat{E}_k}{mk} \left[ \int_{-\infty}^{+\infty} \frac{\partial f_0}{\partial v} \frac{1}{v - \Omega/k} dv + \int_{-\infty}^{+\infty} \frac{\partial f_0}{\partial v} \frac{i\gamma/k}{(v - \Omega/k)^2} dv \right] \\
&= -i \frac{e \hat{E}_k}{mk} \left[ \frac{n_0}{(\Omega/k)^2} - \frac{2 i n_0 \gamma/k}{(\Omega/k)^3} \right] \quad (24)
\end{aligned}$$

where we have neglected the term $(\gamma/k)^2$ in the denominate of the second line since $\gamma \ll \Omega$. $n_1^w$ is contributed from the particles in resonance with the wave. Taking $\frac{\partial f_0}{\partial v}$ to be a constant, we have

$$\begin{aligned}
n_1^w &= -i \int_{-\infty}^{+\infty} \frac{e \hat{E}_k}{mk} \frac{\partial f_0}{\partial v} \left[ \frac{1}{v - \omega/k} + S \Delta(v, \omega/k) \right] dv \\
&= -i \frac{e \hat{E}_k}{mk} \frac{\partial f_0}{\partial v} \bigg|_{v = \frac{\Omega}{k}} \int_{-\infty}^{+\infty} \left[ \frac{v - \Omega/k + i\gamma/k}{(v - \Omega/k)^2 + (\gamma/k)^2} + S \Delta(v, \omega/k) \right] dv \quad (25)
\end{aligned}$$

Changing to the coordinate system travelling with the wave velocity, one obtains

$$\begin{aligned}
n_1^w &= -i \frac{e \hat{E}_k}{mk} \frac{\partial f_0}{\partial v} \bigg|_{v = \frac{\omega}{k}} \int_{-\infty}^{+\infty} \left[ \frac{v + i\gamma/k}{v^2 + (\gamma/k)^2} + S \Delta(v, \omega/k) \right] dv \\
&= -i \frac{e \hat{E}_k}{mk} \left[ i\pi \frac{\partial f_0}{\partial v} \bigg|_{v = \frac{\omega}{k}} - \gamma 2\pi \frac{\partial^2 f_0}{\partial v^2} \bigg|_{v = \frac{\omega}{k}} + \cdots \right] \quad (26)
\end{aligned}$$

Substitution of (24) (26) into the Poisson equation (21) yields respectively from the real and

the imaginary parts

$$\gamma = \frac{\pi \Omega^3}{2n_0 k^2} \frac{\partial f_0}{\partial v}\bigg|_{v=\frac{\omega}{k}} \quad (27)$$

and

$$\Omega = \omega_{pe} = \sqrt{\frac{n_0 e^2}{\epsilon_0 m}} \quad (28)$$

They are exactly the same as those given by Wesson.

In summary, we have presented the general form of solution to Vlasov equation for the electrostatic plasma wave, shown by Eqs. (9) (10) and (17). From this solution, one can get the quasi-linear diffusion coefficients valid for both damping and growing modes. The solution can also lead to the well-known Landau damping and it is equivalent to Wesson's recent explanation of Landau damping using real variables.

ACKNOWLEDGEMENT

This work is supported by National Natural Science Foundation of China under Grant No. 11175213.

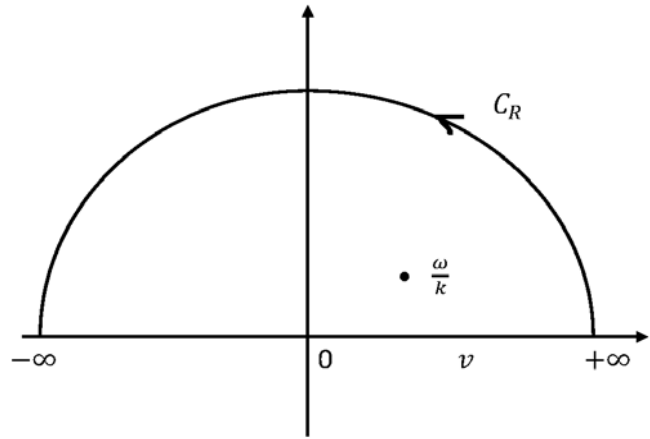

Fig. 1 The integration route used in calculation of density perturbation.